\documentclass[pra,twocolumn,showpacs,aps,floatfix,amsmath,amssymb,amsfonts]{revtex4-1}

\usepackage{amsmath}
\usepackage{graphicx}
\usepackage{color}

\newcommand{\beq}{\begin{equation}}
\newcommand{\eeq}{\end{equation}}

\begin{document}

\title{Nonuniversal beyond mean field properties of quasi-two-dimensional dipolar Bose gases}

\author{Krzysztof Jachymski$^1$}
\author{Rafa{\l} O{\l}dziejewski$^2$}


\affiliation{
$^1$ Institute for Theoretical Physics III, University of Stuttgart, Pfaffenwaldring 57, 70569 Stuttgart, Germany\\
$^2$ Center for Theoretical Physics, Polish Academy of Sciences, Al. Lotnik\'{o}w 32/46, 02-668 Warsaw, Poland
}

\date{\today}

\begin{abstract}
We study a quasi-two dimensional gas of bosonic dipoles, calculating the beyond mean field corrections to the ground state energy and chemical potential neglecting the transverse mode structure. We show that the corrections are sensitive to the high momentum part of the interaction and cannot be expressed solely in terms of the scattering length and the dipole strength. While nonuniversal, the correction is found to be negative, which provides an additional attractive term in the extended Gross-Pitaevskii equation, enhancing the roton instability.
\end{abstract}

\maketitle

\section{Introduction}

Ultracold atomic gases provide a unique opportunity for experimental creation of many-body systems in which the type and strength of interactions as well as the geometry and dimensionality can be controlled very precisely. Using strongly magnetic atoms such as chromium, erbium or dysprosium allows one to access the regime of dipolar interactions~\cite{Lahaye:2007,Lahaye2008,Lu:2011,Aikawa:2012}. This leads to multiple new phenomena already in the weakly interacting regime which can be described by the modified Gross-Pitaevskii equation, such as the magnetostriction effect~\cite{Lahaye:2009,baranov2012condensed}. As the dipolar interaction is partially attractive, it can overcome the short-range repulsion and lead the gas to collapse. Recent experiments~\cite{Kadau2016,Chomaz2016,Schmitt2016} have demonstrated that close to the instability point, the mean field description of the gas is no longer sufficient and the dynamics is governed by the beyond mean field terms. This can lead to a situation in which the mean field attraction is balanced by the quantum fluctuations and the system forms a self-bound liquid quantum droplet~\cite{Petrov2015}. This discovery revived the theoretical interest in calculations of the leading beyond mean field terms~\cite{Lee1957,Beliaev1958,Hugenholtz1959,Lima2012} commonly named the Lee-Huang-Yang (LHY) corrections, which allow for accurate description of the droplet formation~\cite{Barbut2016,Wachtler2016,Blakie2016}.

It is well known that the role of quantum fluctuations is enhanced in lower dimensions. Theoretical description of the two-dimensional Bose gas has shown to be very challenging, and the beyond mean field corrections to the energy were calculated much later than for three- and one-dimensional cases~\cite{Schick1971, Popov1972,Fisher1988,Ovchinnikov1993,Posazhennikova2006,Astrakharchik2007,Astrakharchik2010,Mora2009} and by comparison with Monte Carlo methods have been shown to work only for extremely dilute gases and low temperatures. However, recent works pointed out that quantum droplets similar to the three-dimensional ones can also be formed in two-component mixtures in lower dimensions~\cite{Petrov2016}.
The physics of quasi-two-dimensional quantum dipoles is also very interesting. Confining the gas to two dimensions with dipoles polarized perpendicular to the plane increases the stability of the system as the interactions are dominantly repulsive. However, the excitation spectrum can be nonmonotonic for sufficiently strong dipole and develop a roton minimum, which provides low energy excitations at finite momentum~\cite{Santos2003,Uwe2006,Ronen2007,Sinha2007,Boudjemaa2013,Mishra2016,Chomaz2017,Boudjemaa2017}. This makes the system unstable towards collapse. If some stabilization mechanism existed, the roton instability could lead to the development of a new ground state with spatial order induced by the finite roton momentum, forming a stable supersolid phase~\cite{Boninsegni2012,li2017stripe,leonard2017supersolid}. Such a stabilization can be achieved e.g. by means of three-body repulsion~\cite{Lu2015stable}. 

As the beyond mean field corrections in three dimensions stabilize the dipolar gas against collapse and lead to the droplet formation, derivation of the analogous two-dimensional result is desirable. Motivated by this, in this work we aim to compute the corrections for a simple two-dimensional model with an effective interaction potential.
We estimate the size of beyond mean field corrections and find that they are negative, failing to halt the roton instability. Moreover, we show that one cannot find a universal description of the system and instead obtain cutoff-dependent corrections to the chemical potential, indicating that the dynamics of the realistic gas will strongly depend on the details of the interaction and the external trapping potential.

The paper is organized as follows. In Sec.~II we introduce the model of a quasi-2D dipolar Bose gas and discuss the form of the effective interaction. In Sec.~III, we calculate and discuss the ground state energy and the correction to the chemical potential. Discussion of the role of beyond mean field terms for the system dynamics is provided in Sec.~IV. Conclusions are drawn in Sec.~V.

\section{Model}
The considered system is a dilute gas of dipolar bosons tightly trapped in one direction and free to move in a plane. We disregard the dynamics in the trapped direction and start with the general many-body Hamiltonian describing the two-dimensional Bose gas
\beq\label{Ham}
H = \sum_{\mathbf{k}}\frac{k^2}{2}\hat{a}_{k}^\dagger \hat{a}_{k}+\frac{1}{2}\sum_{\mathbf{k},\mathbf{k^\prime}, \mathbf{q}} \hat{a}_{ \mathbf{k}+\mathbf{q}}^\dagger \hat{a}_{\mathbf{k^\prime}-\mathbf{q}}^\dagger V(\mathbf{q}) \hat{a}_{\mathbf{k}}\hat{a}_{\mathbf{k^\prime}},
\eeq

Here, $\hat{a}$ is the bosonic annihilation operator, $\mathbf{k}$ is the 2D momentum, $V(\mathbf{q})$ is the effective two-dimensional two-body interaction resulting from the interplay of the 3D interaction and confinement and we have set $\hbar$ and the atomic mass $m$ to unity for notational convenience.

As we are interested in the dilute, zero temperature properties, the effective interaction term has only to reproduce the low energy scattering properties. Due to the presence of dipolar interactions, the potential in position space has a long-range $r^{-3}$ tail which in momentum space translates to a term linear in $q$. The scattering amplitude should then follow the generic form~\cite{Boudjemaa2013}
\beq
\label{eq:intsimple}
V_{{\rm eff}}(k)=g-C_{\rm dd} k
\eeq
where $C_{\rm dd}$ describes the strength of the dipolar part of the potential and $g$ depends on the details of the interaction as well as on the external trap confining the system to two dimensions, which leads to a confinement-induced shift to the bare value of $g$~\cite{Olshanii1998,Petrov2001,Sinha2007}. We assume that the planar confinement is strong and that the gas density is very low so that the healing length remains much smaller than transverse trap length scales. Precise calculation of the scattering amplitude in the presence of realistic confinement can be performed numerically. However, at this point we want to extract the features which are independent of the interaction and trap details and express the results in terms of the scattering length and the dipole strength. We thus choose to work with the potential~\eqref{eq:intsimple} with a high momentum cutoff $\kappa$. The limits of this approach will become evident in the next section. 

\section{Calculation of the energy}
\subsection{Momentum cutoff method}
\begin{figure}
\includegraphics[width=0.5\textwidth]{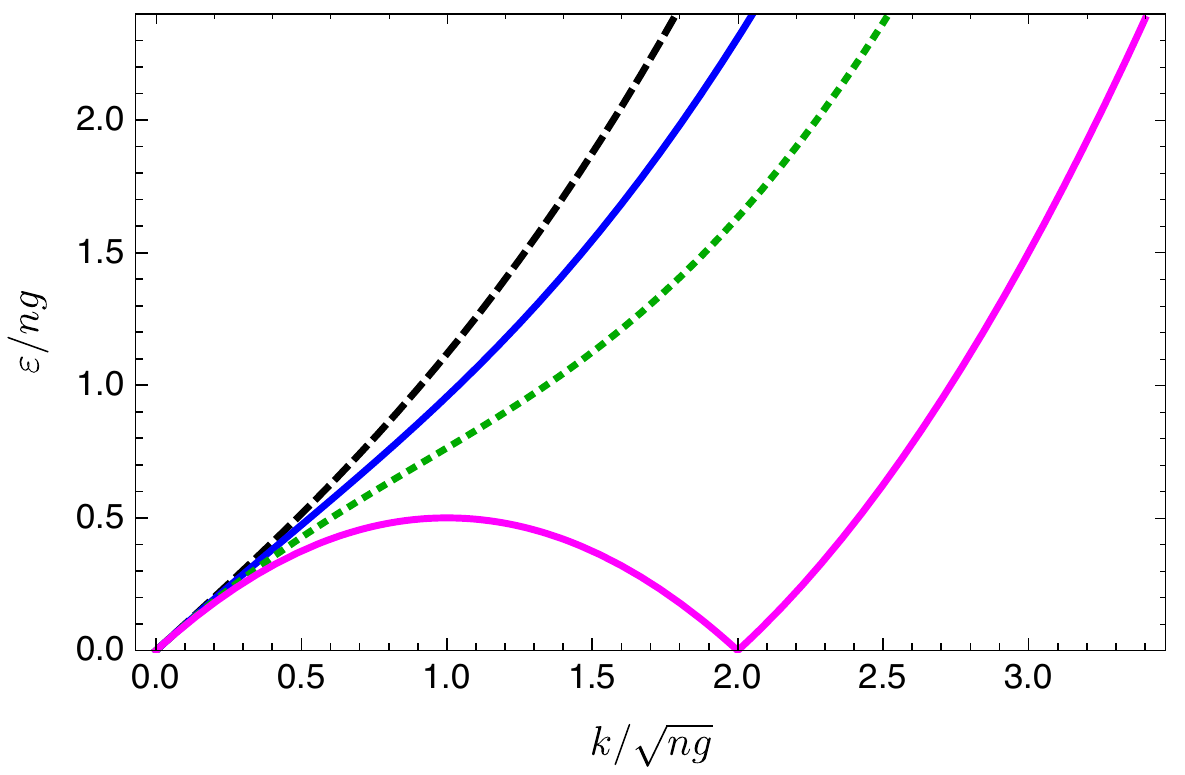}
\caption{\label{fig:bogoliubov}Bogoliubov dispersion for different dipole strength $C_{\rm dd}=0$ (black dashed), $1/3\sqrt{g/n}$ (blue), $2/3\sqrt{g/n}$ (green dotted) and the critical value $\sqrt{g/n}$ (magenta) for which the spectrum has a roton minimum which touches zero.}
\end{figure}

In order to describe the system within the framework of the mean field theory at zero temperature, we proceed with the standard Bogoliubov method (note that for 2D Bose gases the hydrodynamic approach of Popov~\cite{Popov1972} is better suited as corrections beyond the standard LHY term are of similar order; however, we do not aim to achieve this level of accuracy here). In the lowest order we neglect all nonzero momenta in Eq.~\eqref{Ham}. This cancels the dependence on the $C_{\rm dd}$, as it is well known that in the ultra dilute limit the dipolar Bose gas in two dimensions can be described solely in terms of contact interactions. In the second order, we find the standard Bogoliubov Hamiltonian
\beq
H_{\rm eff}=E_0+\sum_{\mathbf{k}}{\varepsilon(\mathbf{k}) \alpha^\dagger_\mathbf{k}\alpha_\mathbf{k}}
\eeq
with the dispersion relation
\beq
\varepsilon(k)=\sqrt{\left(\frac{k^4}{4}+k^2 nV(k)\right)},
\eeq
where $n$ is the 2D gas density. As the interaction can take negative values at finite momenta, the excitations can acquire an imaginary part and cause roton instability. For the interaction given by~Eq.~\eqref{eq:intsimple} this takes place at the critical dipole strength  $C_{\rm dd}=\sqrt{g/n}$. The instability is located at the characteristic momentum $k^\star=2\sqrt{n g}$, which will be important further on. Some examples of the Bogoliubov dispersion are shown in Fig.~\ref{fig:bogoliubov}. 

The ground state energy density $E_0/V$, which is relevant for the equation of state at zero temperature, contains the sum over the zero-point energies of the Bogoliubov modes
\beq
\label{eq:enlhy}
E_0/V=\frac{1}{2}gn^2+\frac{1}{2}\sum_{\mathbf{k}}{\left(\varepsilon_k-k^2/2-n V(k)\right)}.
\eeq

The second part is the analogue of the LHY correction in three dimensions. Calculation of the correction integral over infinite momentum space turns out to be divergent, as is the case also in 3D. The divergence is cured by the cutoff $\kappa$. However, this requires careful treatment, as using the effective potential with a cutoff impacts the resulting scattering length, which is the physical parameter describing the system. This can be seen e.g. in the second order Born approximation, as shown in the next subsection. In order to restore cutoff-independent results, we have to link $\kappa$ with the coupling constant $g$~\cite{Petrov2016} so that the 2D scattering properties expressed using the two-dimensional scattering length $a$, or equivalently $\epsilon=4e^{-2\gamma}/a^2$ with $\gamma$ being the Euler's constant, are preserved. Furthermore, one needs to verify that the chosen cutoff is larger than the typical momenta in the gas set by $\sqrt{ng}$ and in particular larger than the roton momentum $2\sqrt{ng}$, otherwise the approach is not consistent. On the other hand, it will be shown in the next subsection that the cutoff cannot be orders of magnitude larger than $\sqrt{ng}$, as the correction terms will contain expressions of the type~$\ln\left( ng/\kappa^2\right)$, which need to be much smaller than unity. 

Having obtained the energy density of the ground state, the chemical potential can be extracted according to its definition $\mu=\frac{\partial E}{\partial N}$. In the lowest order, we get the standard result $\mu=ng$, with an additional correction from the second term in Eq.~\eqref{eq:enlhy}.

\subsection{Contact interactions}
In order to better understand the properties of 2D systems and limitations of our method, let us first consider the case of contact interactions. In Eq.~\eqref{Ham}, we take
$V_{\rm eff}(\mathbf{k})=g$ for $k<\kappa$ and $0$ otherwise. For this potential we calculate the low energy $t$ matrix making use of the Born approximation up to the second order, obtaining
\beq
t(z)\approx g(1-g \ln\left(-\kappa^2/z\right)/4\pi),
\eeq
where the calculation is done at a small negative energy $z$ to avoid divergencies. By definition, the $t$ matrix should approach the form
\beq
t(z) = 4\pi/\ln\left(-\epsilon/z\right),
\eeq
where as previously $\epsilon=4e^{-2\gamma}/a^2$ and $a$ is the two-dimensional scattering length. By comparing the Born approximation result with the definition in the limit of small $g$ (weak interactions), we obtain $g(\kappa)=4\pi/\ln\left(\epsilon/\kappa^2\right)$. Note that the mean field theory can only be valid when both $\left| t\right|$ and $\left| g\right|$ are small and the system is ultradilute so that the typical energy $|z|\sim n$ is also small and the Born series converge. 

\begin{figure}
\includegraphics[width=0.5\textwidth]{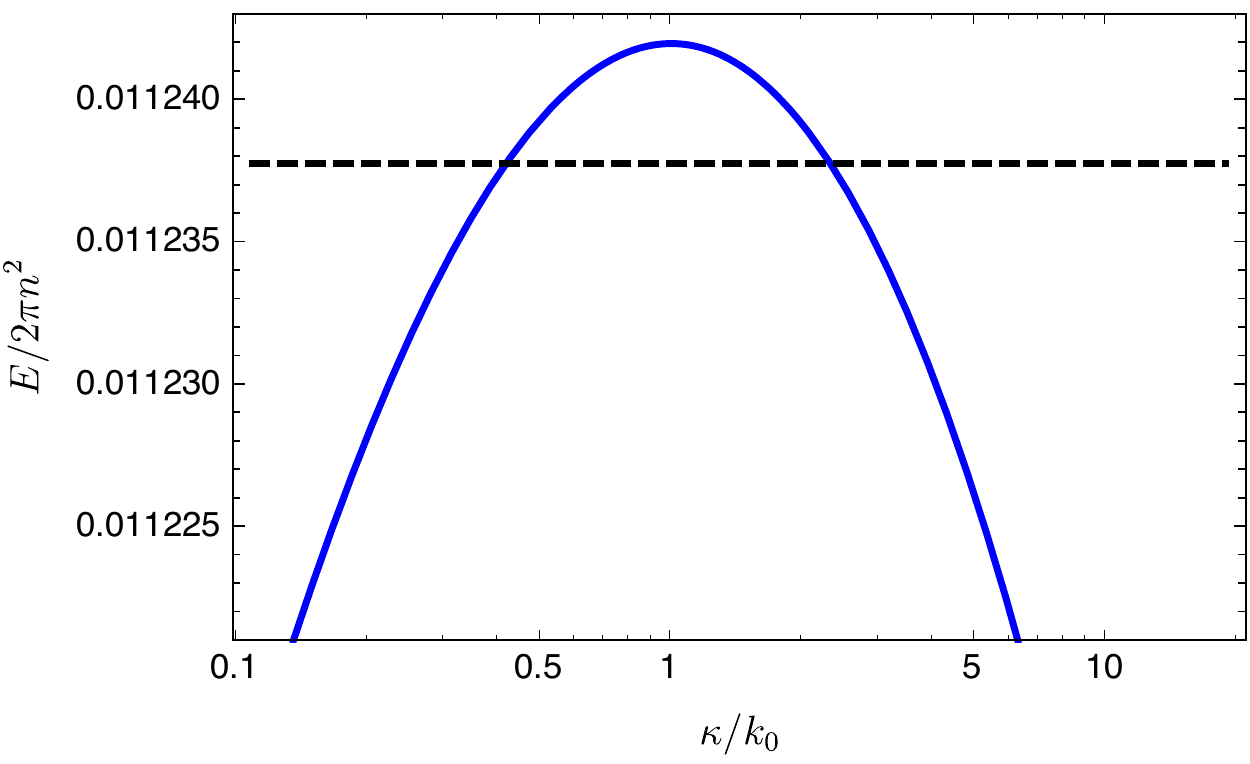}
\caption{\label{fig:contact}Energy of the system as a function of the cutoff parameter $\kappa$ for the density $n=10^{-10}$ and $\epsilon=10^{28}$. The dashed line shows the result given by~\eqref{eq:popov}. Here $k_0=\sqrt{g_0 n}$ and $g_0=4\pi/\ln\left(\epsilon/n\right)$.}
\end{figure}

We now calculate the total energy as a function of $\kappa$. The correction for this case can be calculated analytically. After a series expansion of the energy assuming large $\kappa$, one obtains
\beq
E=\frac{1}{2}g n^2+\frac{g^2 n^2}{8\pi}\ln\left(\frac{\sqrt{e}gn}{\kappa^2}\right),
\eeq
in agreement with the analogous result for a two-component mixture~\cite{Petrov2016}. Plotting the energy as a function of $\kappa$, we find a narrow range in which the results are weakly dependent on the cutoff. The behavior of $E(\kappa)$ is shown in Fig.~\ref{fig:contact}. Here and further on we use $g_0=4\pi/\ln\left(\epsilon/n\right)$ and $k_0=\sqrt{n g_0}$ to set the scale.  Choosing the value of $\kappa$ at the local maximum, we obtain results which are in good agreement with the analytic result~\cite{Popov1972,Astrakharchik2010}
\beq
\label{eq:popov}
\frac{E}{V}=\frac{2\pi n^2}{\left|\ln na^2\right|+\ln\left|\ln na^2\right|+C_1+\frac{\ln\left|\ln na^2\right|+C_2}{\left|\ln na^2\right|}},
\eeq
where the coefficients $C_1\approx -2.8$ and $C_2\approx -0.05$. We observe that the optimal range of $\kappa$ values agrees with the requirement of being larger but not exponentially larger than $\sqrt{ng}$.

\subsection{Dipolar interactions}
Let us now discuss how the method performs when applied to the dipolar interaction case.  
We take $V(\mathbf{k})= g-C_{\rm dd}k$ for $k<\kappa$ and zero otherwise. The low energy $t$ matrix corresponding to this potential within the second order Born approximation is
\beq
t=g- C_{\rm dd}\sqrt{-z}+(g^2+C_{\rm dd}^2 z) \ln\left(\frac{-z}{\kappa ^2}\right)+\frac{\kappa  C_{\rm dd}}{4\pi} (4 g-\kappa  C_{\rm dd}).
\eeq 
We note that the second order term does not impact the part linear in momentum, so renormalization of $g$ can be expected to still give good results. On the other hand, the dipolar term contributes to the scattering length value. With the last term being of the order of $g^2$, we proceed as before and take $g=4\pi/\ln\left(\epsilon/\kappa^2\right)$. 

\begin{figure}
\includegraphics[width=0.5\textwidth]{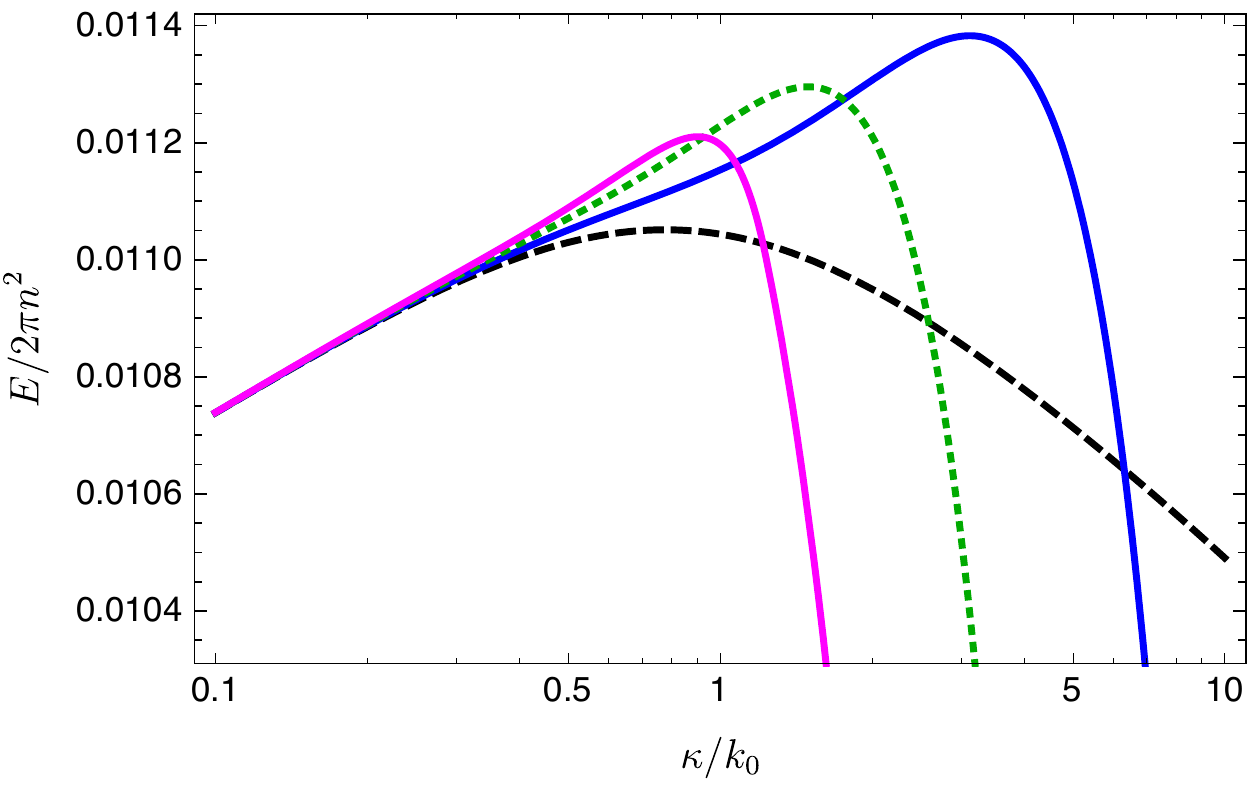}
\caption{\label{fig:en_dip}Energy of the system as a function of the cutoff parameter $\kappa$ for different dipole strengths $C_{\rm dd}=0$ (black dashed), $1/3\sqrt{g_0/n_0}$ (blue), $2/3\sqrt{g_0/n_0}$ (green dotted) and $\sqrt{g_0/n_0}$ (magenta). Density and $\epsilon$ are the same as in Fig.~\ref{fig:contact}.}
\end{figure}

\begin{figure}
\includegraphics[width=0.5\textwidth]{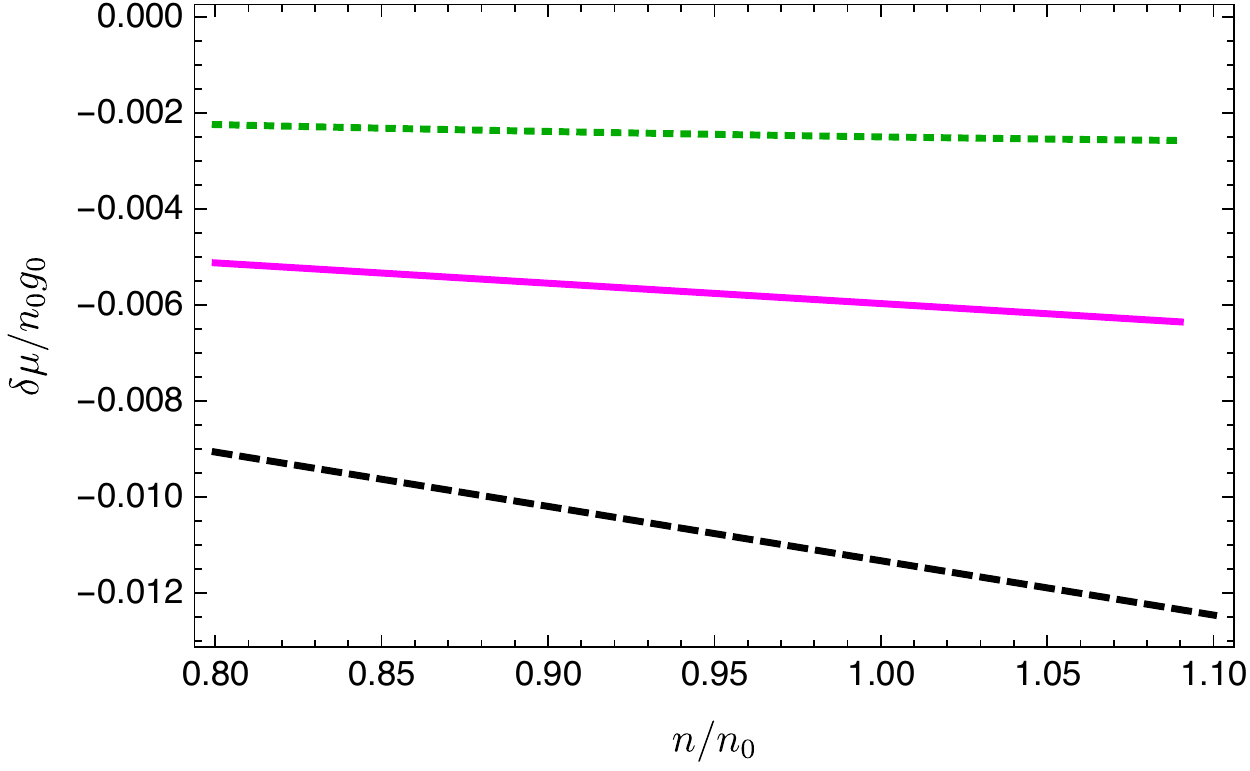}
\caption{\label{fig:mu_n}Correction to the chemical potential as a function of the density for various dipole strengths $C_{\rm dd}=0$ (black dashed), $2/3\sqrt{g_0/n_0}$ (green dotted) and $\sqrt{g_0/n_0}$ (magenta). The cutoff at each point is chosen to minimize the derivative $\frac{\partial E}{\partial \kappa}$.}
\end{figure}

We now perform the integration over the Bogoliubov spectrum to calculate the correction to the energy and obtain a complicated expression with terms linear and quadratic in $\kappa$ in addition to the ones similar to the contact interaction
\beq
\delta E=n^2\frac{\kappa C_{\rm dd} \left(4 g-\kappa  C_{\rm dd}-4 n C_{\rm dd}^2\right)}{8 \pi}+o(\kappa).
\eeq
There are many more terms of higher order, their precise form is not relevant. Let us estimate the importance of this term close to the roton conditions. We take $g\sim g_0$, $\kappa\sim \sqrt{ng_0}$ and $C_{\rm dd}\sim\sqrt{g_0/n}$. This leads to $\delta E \propto \left(n g_0\right)^2$, which looks like a proper perturbative result since $ng_0$ is a small parameter. 
The crucial problem in describing the properties of the system in the presence of the roton is that the roton minimum is located at momenta $k\sim 2\sqrt{ng}$, and the cutoff we find lies in the same range. Specifically, as the dipole strength approaches the critical value, the maximum of energy shifts towards values of $\kappa$ which are smaller than $k^\star$ and the calculation does not take into account the contribution from the unstable region. This behavior is shown in Fig.~\ref{fig:en_dip}. In the following, we take $\epsilon=10^{28}$ and the density scale $n_0=10^{-10}$ to demonstrate the results, and express the dipole strength in units of $\sqrt{g_0/n_0}$. The calculated corrections to the chemical potential as a function of the density are presented in Fig.~\ref{fig:mu_n}. As the cutoff does not cover the roton, there is no spectacular change of the behavior at the critical density $n_0$ for strong dipoles. Instead, the correction stays rather flat.

\begin{figure}
\includegraphics[width=0.5\textwidth]{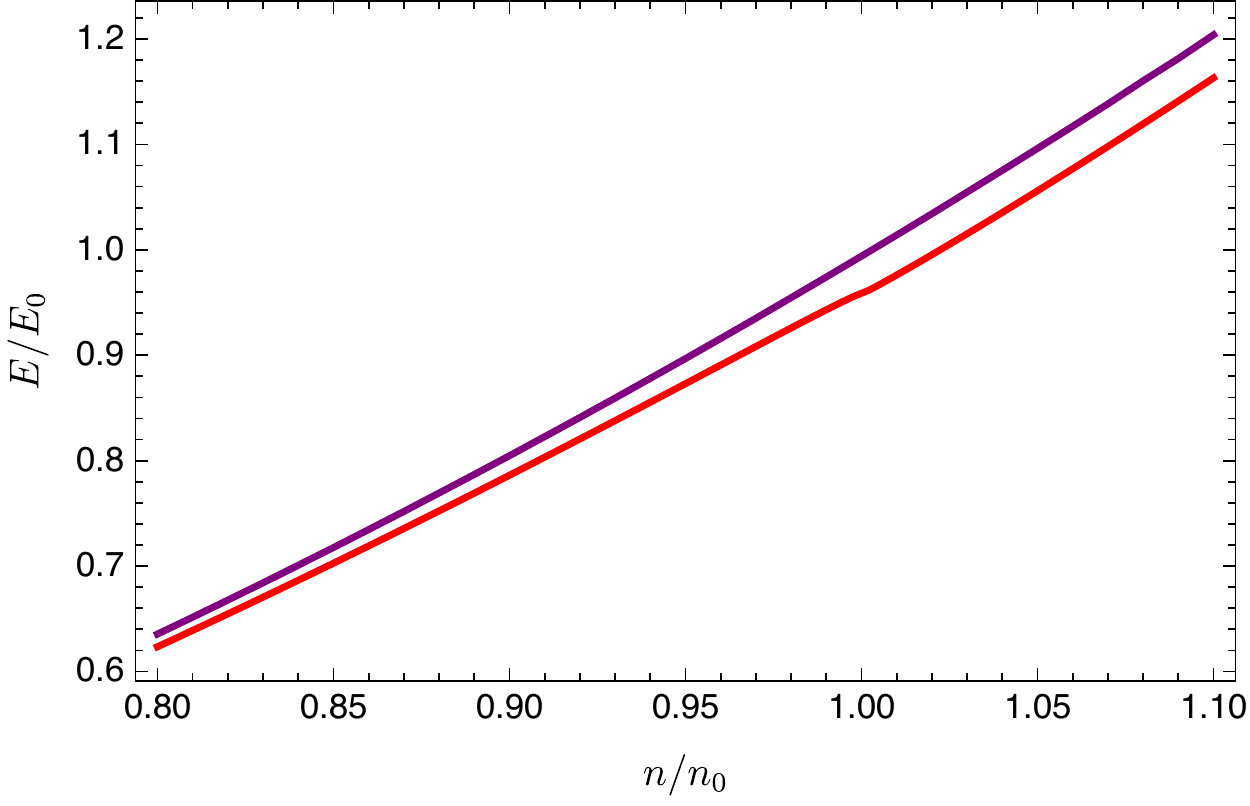}
\caption{\label{fig:enroton}Ground state energy of the gas as a function of the density in the vicinity of the roton with $C_{\rm dd}=\sqrt{g_0/n_0}$. The purple line shows the result with the cutoff minimizing the derivative $\frac{\partial E}{\partial \kappa}$, while for the red line the $\kappa$ value was shifted to $1.06\,k^\star$ to cover the roton.}
\end{figure}

\begin{figure}
\includegraphics[width=0.5\textwidth]{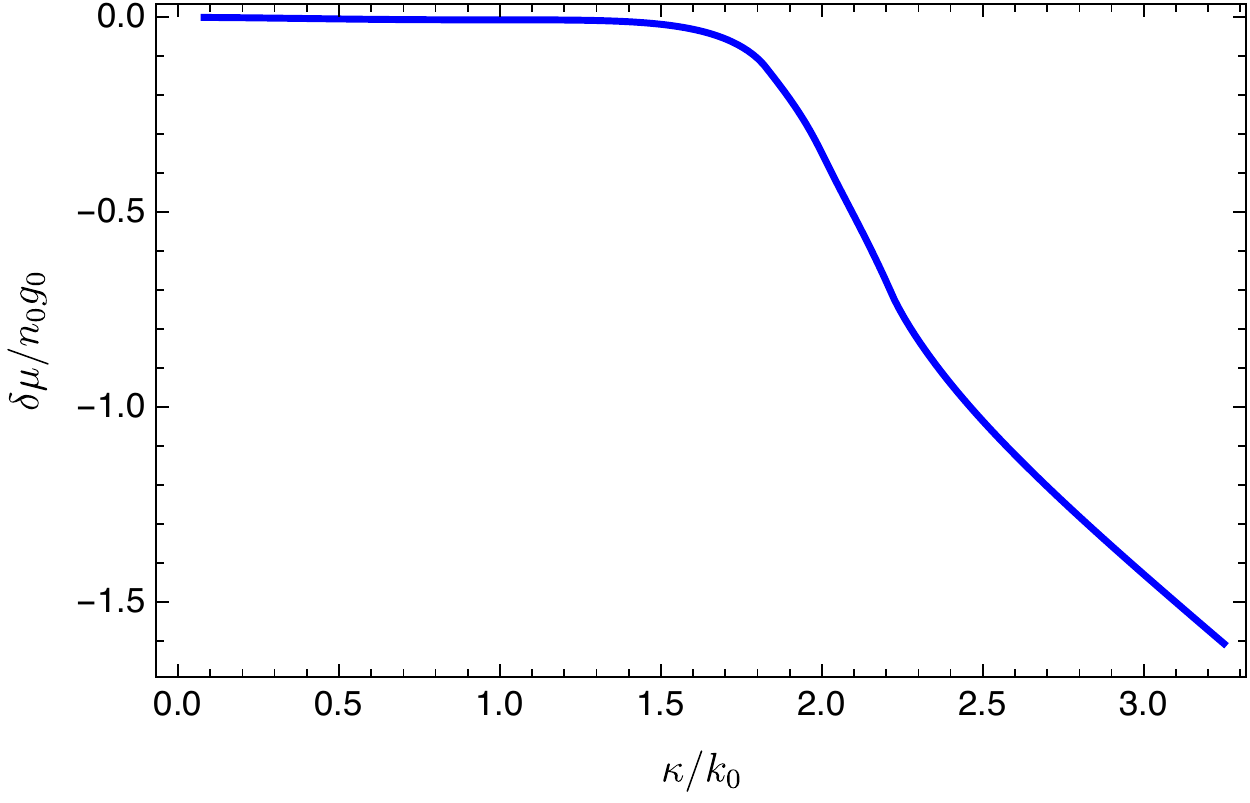}
\caption{\label{fig:mukappa}Correction to the chemical potential for strong dipolar interaction $C_{\rm dd}=\sqrt{g_0/n_0}$ at the critical density $n_0$ as a function of the cutoff $\kappa$.}
\end{figure}

As long as the energy is not strongly affected, we can slightly shift the cutoff from the maximum towards larger values and probe the roton region. As shown in Fig.~\ref{fig:enroton}, shifting the cutoff value to $1.06\,k^\star$ introduces a relatively small change in the ground state energy. At the same time, the behavior of the correction to the chemical potential is dramatically altered (see Fig.~\ref{fig:mu_n}). Firstly, its magnitude becomes an order of magnitude larger. More importantly, its qualitative behavior also changes as it starts to diverge at the critical density ($n_0$ for $C_{\rm dd}=\sqrt{g_0/n_0}$). This shows that within our method, the quasi-two dimensional dipolar Bose gas is nonuniversal, i.e. the properties of the system strongly depend on the high momentum cutoff. This is further illustrated in Fig.~\ref{fig:mukappa} which shows the dependence of $\delta\mu$ on $\kappa$ at constant density $n_0$ and fixed dipole strength $C_{\rm dd}=\sqrt{g_0/n_0}$. In the region close to $k^\star=2k_0$, the correction varies rapidly.

\begin{figure}
\includegraphics[width=0.5\textwidth]{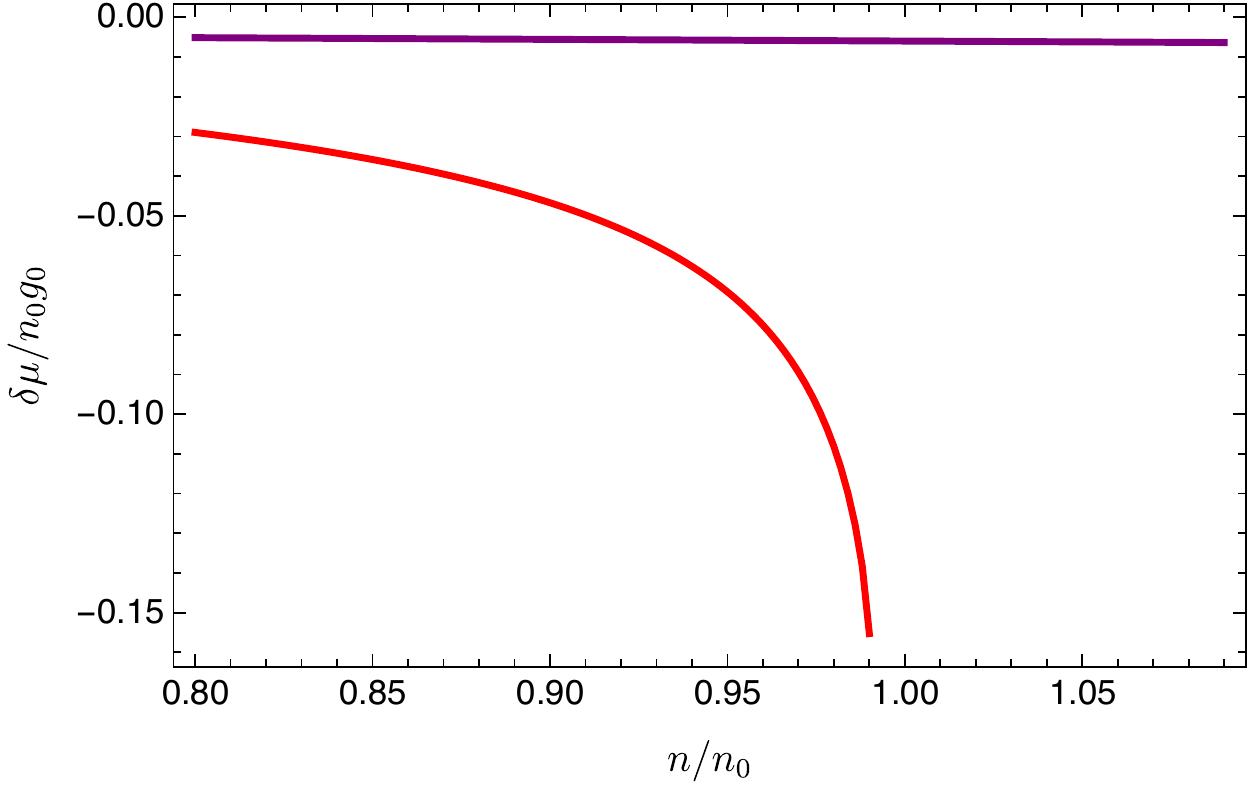}
\caption{\label{fig:muroton}Correction to the chemical potential as a function of the density for strong dipolar interaction $C_{\rm dd}=\sqrt{g_0/n_0}$ and the cutoff chosen as in Fig.~\ref{fig:enroton}.}
\end{figure}

Interestingly, these results suggest that one can expect strong dependence of the properties of the gas on the strength of the transverse confinement. The first effect that the trap can induce is the well known confinement-induced resonance phenomenon~\cite{Olshanii1998,Petrov2001} in which the value of $g$ is modified. Here we expect more intriguing effects. 
The effective quasi-2D interaction potential for two dipoles in a pancake shape trap can be estimated as an integral of the full 3D dipolar interaction over the transverse harmonic oscillator ground state with characteristic width $l_\perp$. This results in
\beq
\label{effpot}
V_{{\rm eff}}(k)=g-C_{\rm dd}kl_{\perp}\exp{\left(k^2l_{\perp}^2/2\right)}{\rm {Erfc}}(kl_{\perp}/\sqrt{2}) \, .
\eeq
Here ${\rm Erfc}$ denotes the complementary error function. This potential is not accurate especially at high momenta, as the wave function at short interparticle distances can be expected to deviate very strongly from the transverse trap ground state, but serves as a first order approximation. The series expansion of this potential at small momenta reads $V_{{\rm eff}}(k)=g-C_{\rm dd}l_\perp k+o(k^2)$. Here $l_\perp$ sets the scale for the strength of the linear term. At $k\gg 1/l_\perp$ the potential approaches a constant value. Manipulating the $l_\perp$ can then allow to tune the position of the crossover regime, which would correspond to shifting the cutoff in our model. More elaborate calculations have to be performed to verify this conjecture.

\section{Role of the corrections for beyond mean field dynamics}
The important role of the LHY correction for the dynamics of dipolar gases as well as two-component mixtures at the stability boundary is now well established~\cite{Petrov2015,Barbut2016,Chomaz2016}. Let us now estimate the impact of the corrections on the two-dimensional dipolar gas. In order to do this, we consider the extended Gross-Pitaevski equation in which the correction to the chemical potential $\delta\mu$ is incorporated as an additional density-dependent term, making use of the local density approximation~\cite{Petrov2015}
   \begin{equation}\label{eq:eGPE}
    i\hbar \partial_t \psi(x,y) = \bigg[ -\frac{\nabla^2}{2}  + \mathcal{F}^{-1}\bigg[ V_{{\rm eff}}({\bf k}) \bigg] + \delta\mu \bigg] \psi(x,y)
    \end{equation}
where $\mathcal{F}^{-1}$ stands for the inverse Fourier transform. As shown in previous paragraphs, the LHY correction $\delta\mu$ is negative in our model and provides additional attraction. As a result, the correction supports the collapse of the atomic cloud near the roton instability, making a small shift to the stability boundary towards weaker dipoles.

In three dimensions, it is possible to reduce the mean field energy of the gas to zero by balancing the dipolar and contact interaction contribution. This leads to the formation of finite size droplets close to the stability boundary. However, in our case the droplets cannot be created as the ground state energy does not have a minimum at finite density. This stems from the fact that the mean field term $\propto n^2 g$ is not cancelled by any other term and the corrections cannot take the leading role. Tilting the dipoles and introducing partial attraction in the 2D system would change this behavior.

Finally, two-dimensional ultracold dipolar gas can be considered as a candidate for creation of the supersolid phase, in which the system displays crystalline order while being superfluid at the same time. As shown by~\cite{Lu2015stable}, on the mean field level this requires very specific conditions including additional three-body repulsive interaction to stabilize the system. We conducted a similar analysis including the quantum fluctuations in the energy functional based on the Gross-Pitaevskii equation and testing different variational wave function. We find that the $g<0$ case is always unstable in the absence of three-body repulsion, as the beyond mean field correction is negative. For $g>0$ the only stable solution is given by the uniform state, which means that the supersolid phase cannot be supported. However, these estimations may be modified by including the high momentum part of the LHY correction coming from the transverse modes.

\section{Conclusions}
We have analyzed the beyond mean field corrections to the equation of state of a quasi-two-dimensional gas of dipolar bosons using a simplistic model in which the high momentum part of the interaction is described using the cutoff. Obtaining universal results close to the region of the roton instability turned out to be unfeasible. The presence of terms linear in momentum in the Hamiltonian introduces a challenge for the theoretical description, as the methods available for treating quadratic effective range corrections~\cite{Salasnich2017} are not directly applicable.
Nevertheless, we have found that the correction to the chemical potential is negative, which results in an effective attraction in the extended Gross-Pitaevskii equation.

It is important to stress that in this calculation the impact of transverse trap modes has been neglected. While this approximation holds at the level of the Gross-Pitaevskii equation for sufficiently tight traps, these modes can strongly impact the correction term~\cite{Edler2017,Zin2018,Ilg2018}. Our results suggest that the quasi-2D dipolar Bose gas cannot be precisely described using the effective two-dimensional theory, as there are potentially important high-momentum contributions which can even provide some stabilization mechanism. Formulation of an extended model with more realistic high-momentum part will be the subject of our future work.


In order to obtain deeper insight into the problem, it would be desirable to perform more involved calculations. Here one idea would be to consider the full three-dimensional model in the presence of anisotropic trap and dipolar interaction and numerically diagonalize the Bogoliubov-de Gennes equations, going beyond the local density approximation~\cite{Ticknor2012}. Obtaining reliable, numerically stable correction for this case would require extensive numerical effort as well as careful linking to the two-dimensional quantities. 

Another promising direction is to consider the case of tilted dipoles, in which the interaction is no longer isotropic in space, resulting in a striped phase which can provide the route to supersolidity~\cite{Wenzel2017,Baillie2017,Bombin2017}. It may be possible to engineer the system parameters such that the LHY term would enhance the stability of the system even for the dilute case.

We would like to thank Tobias Ilg, Tomasz Wasak, Hans Peter B\"{u}chler and especially Dmitry Petrov for inspiring discussions and useful remarks. This work was supported by Alexander von Humboldt Foundation (K.J.) and the (Polish) National Science Center Grant 2015/19/B/ST2/02820 (R.O.).

\bibliography{Allrefs}
\end{document}